\documentclass[12pt,preprint]{aastex}

\shorttitle{Magnetic Fields at the Base of the CZ}
\shortauthors{Chou \& Serebryanskiy}

\begin{document}

\title{Searching for the Signature of the Magnetic Fields at the Base of the 
 Solar Convection Zone with Solar-Cycle Variations of p-Mode Travel Time}

\author{Dean-Yi Chou and Alexander Serebryanskiy}

\affil{Institute of Astronomy and Department of Physics,
Tsing Hua University, Hsinchu, 30043, Taiwan, R.O.C.}

\begin{abstract}
We study the solar-cycle variations of solar p-mode travel time for 
different wave packets
to probe the magnetic fields at the base of the solar convection zone. 
We select the wave packets which return to the same spatial 
point after traveling around the Sun with integral number of bounces. 
The change in one-bounce travel time at solar maximum relative to minimum 
is approximately the 
same for all wave packets studied except a wave packet whose lower
turning point is located at the base of the convection zone.
This particular wave packet has an additional decrease in travel time 
at solar maximum relative to other wave packets.  
The magnitude of the additional decrease in travel time 
for this particular wave packet increases with solar activity.  
This additional decrease in travel time might be caused 
by the magnetic field perturbation
and sound speed perturbation at the base of the convection zone.
With the assumption that this additional decrease is caused only by the 
magnetic field perturbation at the base of the convection zone,  
the field strength is estimated to be about $4-7\times 10^{5}$ gauss 
at solar maximum if the filling factor is unity. 
We also discuss the problem of this interpretation.
\end{abstract}

\keywords{Sun: magnetic fields --- Sun: helioseismology --- Sun: interior --- 
 Sun: evolution}

\section{Introduction}

Observations give evidence that magnetic fields on the Sun emerge from below.
How and where magnetic fields are generated is a long standing 
unanswered question in astronomy \cite{cow34,par55,bab61}.  
It has been suggested that the boundary between the radiative zone
and convection zone (CZ) is the best location for an oscillatory solar dynamo
\cite{spi80,par93,cha97}.
Many attempts have been made to detect the magnetic fields in this region
\cite{gou96,bas97,how99,bas00,bas01,eff01,ant01}.
Until now no clear evidence of magnetic field in this region
has been found.  Here we use the technique of time-distance analysis 
\cite{duv93} to measure 
solar cycle variations of travel time of acoustic waves
with different ray paths to probe the magnetic fields at the base of the CZ.

A resonant solar p-mode is trapped and multiply reflected in a cavity between
the surface and a layer in the solar interior.  
The acoustic signal emanating from a point at the surface 
propagates downward to the
bottom of the cavity and back to the surface at a different horizontal
distance from the original point.  Different p-modes have different paths
and arrive at the surface with different travel times and different distances
from the original point. 
The modes with the same angular phase velocity have approximately the same
ray path and form a wave packet.  
The relation between the travel time and travel distance of a wave packet can 
be measured by using the temporal cross-correlation
between the time series at two points \cite{duv93}. 

Different wave packets penetrate into different depths:
the wave packet with a larger phase velocity penetrates into a greater depth.
Time-distance analysis measures the travel times of different wave packets
to probe the interior of the Sun at different depths \cite{duv96,kos96,kos00}.
The ray path of wave packet computed from a standard solar model 
with the ray theory for three different phase velocities is shown in Figure 1. 
If a magnetic field is present at the base of the CZ,
it has different effects on different wave packets.
It changes the travel time of wave packets which can penetrate
into the base of the CZ,
while it has no effect on the wave packets which can not reach 
the base of the CZ.
If the magnetic fields at the base of the CZ vary with the solar cycle
like the surface magnetic fields, 
travel time is expected to vary with the solar cycle as well. 
The change in travel time due to the magnetic fields at the base of the CZ
is small because the ratio of magnetic pressure to gas pressure is small. 
However, the change in travel time increases linearly with the number of 
bounces between the boundaries of the cavity. 
Thus the strategy is to measure the change in multiple-bounce travel time.
Here we measure the time for a wave packet to travel around the Sun 
to come back to the same spatial point.  
If a wave packet takes $N$ bounces
to travel around the Sun, the change in travel time would increase
by a factor of $N$ relative to the change in one-bounce travel time.
Therefore, the problem becomes measuring solar cycle variations of travel time 
with the auto-correlation function of the time series at the same spatial point.
In this study we use two different approaches, the multiple-bounce
travel time analysis (MBTTA) and the power spectrum simulation analysis
(PSSA), to measure the travel time of wave packets. 

\section{Multiple-Bounce Travel Time Analysis (MBTTA)}

In the first approach (MBTTA), we use the helioseismic data taken with the 
Michelson Doppler Imager (MDI) on board the SOHO spacecraft \cite{sch95}.
The MDI data used here are full-disc low-resolution Doppler images 
of $192 \times 192$ pixels,
sampled at a rate of one image per minute.
We have analyzed the data in the period of solar minimum and maximum.
The data are divided into 4096-minute segments.
Each time series of 4096 images is analyzed separately.
The procedure of data analysis is described below. 
(1) Each full-disk Doppler image is transformed into $\sin\theta-\phi$
coordinates, where $\theta$ and $\phi$ are the latitude and the longitude,
respectively, in a spherical coordinate system aligned along the solar 
rotation axis.  
(2) The differential rotation of the solar surface is removed
with an observed surface differential rotation velocity \cite{lib91}.
(3) The data are filtered with a Gaussian filter of FWHM = 2 mHz centered 
at 3 mHz. 
(4) To reduce the interference among the wave packets of different phase
velocities, 
a phase-velocity filter is applied to isolate the signals in a
range of the phase velocity $\omega/[l(l+1)]^{1/2}$, where $\omega$ is
the mode angular frequency and $l$ is the spherical harmonic degree.
For each time series of 4096 images, the signals in the $(\sin\theta,\phi, t)$ 
domain are transformed into the $(l,m,\omega)$ domain, where $m$ is 
the azimuthal order.
A filter is applied to select the signals in a range of $\omega/[l(l+1)]^{1/2}$.
The center of filter is chosen such that the one-bounce travel 
distance is $360^\circ/N$, where $N$ is an integer.
The filtered signals are transformed back to the $(\sin\theta,\phi, t)$ 
domain.
The points inside the active regions are excluded to avoid the contaminated
signals measured in the magnetic regions.
(5) The auto-correlation function of time series at each $(\sin\theta,\phi)$ 
is computed.  
(6) The auto-correlation functions are then averaged over a region of 
$84^\circ \times 94^\circ$ at the disk center.  The size of averaging area
is selected to avoid the signals near the limb.  
(7) The phase travel time $\tau_N$ is determined from the instantaneous phase
of the auto-correlation function with a Hilbert transform technique 
\cite{bra86,duv96}.
It is repeated for $N$ between 6 and 15
to obtain the travel times of different wave packets. 
Since the waves used to compute the auto-correlation begin at a point at 
the surface, propagate in all directions to go around the Sun and
come back to the same point,
the auto-correlation function consists of signals which pass through
a medium element from opposite directions.  
Thus the effects of motion of this element cancel out in the 
auto-correlation function.

To study solar cycle variations,
the travel time $\tau_N$ is averaged over a solar minimum period
(May 1996 - May 1997) and a solar maximum period 
(January 2000 - February 2001). 
It is found that the the travel time at solar maximum is shorter than that at
solar minimum.  The difference is defined as $\delta\tau_N$. 
The change in one-bounce travel time is equal to $\delta\tau_N/N$.
To see the change of different wave packets, we plot 
$\delta\tau_N/N$ versus $N$ in the left panel of Figure 2. 
The result is denoted by the open circles.
The value of $\delta\tau_N/N$ is approximately the same for all $N$'s, 
except a small drop at $N=8$.
This approximately constant change is caused by the magnetic 
fields near the surface because the rays paths of different wave packets near 
the surface are approximately the same.
This change in travel time, corresponding to a fraction of change in travel
time of about $-1.3\times 10^{-4}$, is consistent with the well-known 
solar-cycle variations of frequency, about 0.4 $\mu$Hz for modes at 3 mHz
\cite{lib90}.

The interesting phenomenon in Figure 2 is the additional 
decrease in travel time at $N=8$.
The average of $\delta\tau_N/N$ over all $N$'s except $N=8$ is
$-0.763\pm 0.005$ second. 
The difference between $\delta\tau_8/8$ and the average value
is $0.053\pm 0.022$ second. 

\section{Power Spectrum Simulation Analysis (PSSA)}

The correlation function, from which the travel time is determined,  
is the inverse Fourier transform of the power spectrum of p-modes.  
Thus the signal corresponding to the travel time perturbation detected 
above should also exist in the mode frequencies which are determined 
from the power spectrum.  
In the second approach (PSSA),
we use the mode frequencies measured with the helioseismic data 
taken with MDI and the Global Oscillation 
Network Group (GONG) \cite{har96} to construct the power spectrum
in ($l,\omega$).
The MDI data covers May 1996 - June 2001, and the GONG data May 1995 
- May 2001. 
The mode frequencies are averaged over about one year for MDI and 
about two years for GONG to increase the signal-to-noise ratio.
The range of each averaging period is indicated by the horizontal bar
in Figure 3. 
To construct the power spectrum in ($l,\omega$) 
from the mode frequencies, we assume that the spectral profile of each mode 
is a Lorentzian, and the width is a function of frequency only.
We adopt the width measured in Howe et al. (1999).
The power distribution versus $l$ is taken from the $l-\omega$ diagram
obtained in MBTTA.
The frequency variation of power is obtained by applying a Gaussian 
filter centered at 3 mHz with FWHM = 0.55 mHz.
Power is set zero for frequency greater than 3.5 mHz or 
smaller 2.5 mHz because some of mode frequencies above 3.5 mHz 
are not available.  
The composite power spectrum is filtered with the same phase velocity filter 
as in MBTTA prior to computing the cross-correlation function with the 
inverse Fourier transform. 
The cross-correlation function at zero travel distance corresponds to
the auto-correlation function averaged over all spatial points, which
is used to determine the travel time of the wave packet.
The value of $\delta\tau_N/N$ for various periods along the solar cycle 
is shown in Figure 2.

The results of PSSA also show that there exists an additional decrease 
in travel time at $N=8$ relative to other $N$'s.
The magnitude of additional decrease increases with solar activity.
To quantify the additional decrease at $N=8$, we define $\Delta\tau_8$
as the difference between $\delta\tau_8/8$ and the value at $N=8$ 
determined from a linear fit to $\delta\tau_N/N$ of all other $N$'s.
Figure 3 shows $-\Delta\tau_8$ and the sunspot number versus date.
The results of MDI and GONG are consistent, though the results of GONG
are noisier.
At solar maximum, $-\Delta\tau_8 \approx 0.015$ second  
which is smaller than that from MBTTA.
This difference may be caused by the differences in data analysis. 
First, the width of the frequency filter used in PSSA  
is smaller than that in MBTTA because some of higher 
mode frequencies are not available.  
A narrower filter yields a wider wave packet.
Second, the correlation functions are averaged over a central region of 
$84^\circ \times 94^\circ$ in MBTTA, while the averaging area in PSSA 
is equivalent to the area used to measure
the mode frequencies, which is almost the entire disk.  
The signals near the limb have a lower signal-to-noise ratio. 
Third, the active regions are excluded in MBTTA, while they are not in PSSA. 

\section{Discussion}

To test whether the shorter travel time at $N=8$ is caused by the
analysis procedure, we did the following test.  
The frequency difference between solar maximum and minimum is
smoothed by a fit in the ($l, \nu$) domain.  
The mode frequencies at solar maximum are simulated by adding this 
smooth function to the mode frequencies at minimum.   
Applying the same procedure as in PSSA to the measured frequencies
at solar minimum and the simulated frequencies at solar maximum, 
we compute the change in travel time for different $N$'s.  
The result shows that the change in travel time is approximately 
the same for all $N$'s and there is no additional decrease at $N=8$.
This test indicates that the shorter travel time at $N=8$ is not caused  
by the analysis procedure.

It is unlikely that the shorter travel time at $N=8$ is caused by the spatial 
distribution of the near-surface magnetic fields.
The most prominent spatial pattern of magnetic activity on the surface 
is the active latitudinal band in each hemisphere.  
The separation between the centroids of two active bands is 
about $42^\circ$ in 1998 and monotonically decreases to about $29^\circ$ 
in 2000 (from Greenwich Sunspot Data). 
If the latitudinal distribution of active regions can
cause an additional decrease in one-bounce travel time,
it would occur at $N=9$ in 1998 and shift to $N=12$ in 2000.  
This contradicts to the PSSA results shown in Figure 2.
Thus it is unlikely that the shorter travel time at $N=8$
is caused by the separation of two active latitudinal bands on the surface. 
Since the active longitudes is less prominent than the active 
latitudes, it is unlikely the anomaly at $N=8$ is caused by the
active longitudes. 

The fact that the ray path of the wave packet of $N=8$ has the 
lower turning point 
at the base of the CZ as shown in Figure 1 suggests that the additional 
decrease in travel time at $N=8$ may be caused by the solar-cycle 
varying wave speed at the base of the CZ.
The change in wave speed at the base of the CZ could be caused by 
magnetic field perturbation or/and sound speed ($[\gamma p/\rho]^{1/2}$) 
perturbation.
It has been shown that the global measurements can not distinguish 
these two effects \cite{zwe95}.
The previous study \cite{kos97} has also indicated that
either magnetic field perturbation or sound speed perturbation
alone would cause a change in travel time not only for $N=8$
but also for $N<8$, though it is smaller.
Thus either magnetic field perturbation or sound speed perturbation
alone can not explain the measurements of travel time variation 
shown in Figure 2.  However, the combination of these two perturbations
might be able to explain the measured travel time variations.  
Although we do not know the mechanism producing the sound speed perturbation, 
it probably has the magnetic origin because the presence of a magnetic 
field could change the thermal structure and leads to a change in sound speed.  

With the above caution in mind, 
we will estimate the field strength based on the assumption that 
the additional decrease in travel time at $N=8$
is caused only by the magnetic fields at the base of the CZ.
If we adopt the value of $\Delta\tau_8$ measured with MBTTA and PSSA, 
the fraction of change in travel time due to the wave speed perturbation 
at the base of the CZ is about 
$0.015/\tau_8-0.053/\tau_8 \approx 2.6-9\times 10^{-6}$.
If we use the half width of the tachocline, $0.025 R_{\odot}$, as 
the width of the magnetic layer at the base of the CZ \cite{cor01}, 
the fraction of 
change in wave speed at the base of the CZ, $\delta w /w$, is about 
$2.6-9\times 10^{-5}$ because the wave packet of $N=8$ spends about 
one tenth of time inside the tachocline. 
If the change in wave speed is entirely due to the presence of 
magnetic fields, the fraction of change in wave speed is 
$\delta w/w = \sin^2\theta(v^2_A/c^2)/2$, 
where $v_A=B/(4\pi\rho)^{1/2}$ is the Alfven speed,
and $\theta$ is the angle between wave propagation direction and magnetic 
field \cite{kos00}.
The density $\rho\approx 0.2$ g\,cm$^{-3}$ in the tachocline. 
Averaging $\sin^2\theta$ over all directions yields about 1/2.  
Thus the magnetic field strength  
$B \sim [16\pi \rho c^2 (\delta w/w)]^{1/2} \sim 4-7\times 10^5$ gauss
if the filling factor of magnetic field is unity.
A smaller filling factor would increase the estimated field strength.
The field strength estimated here is greater than most theories 
predict \cite{fis00}.
Such a strong field needs a large degree of subadiabaticity
in the tachocline to stabilize it \cite{gil00}.
The approximately constant $\delta\tau_N/N$ at $N\ge 9$ suggests that 
there is no strong magnetic field in the middle of the CZ.

The problem of above interpretation is that 
no additional decrease in travel time is detected for the neighboring
wave packets of $N=8$ in our measurements. 
The neighboring wave packets, $N=7$ and 9, may be also influenced 
by the magnetic fields at the base of the CZ. 
The influence depends on the width and location of the magnetic layer 
at the base of the CZ and the width of the wave packets.   
If we adopt the parameters used above and the ray approximation 
to estimate the influence on the neighboring wave packets, 
the additional decrease at $N=7$ is about $40\%$ of that at $N=8$, while
there is no effect on $N=9$.
However, the finite width of the wave packets would increase 
the effect of magnetic fields on the neighboring wave packets.  
The previous studies have shown that the travel-time sensitivity kernel is wide 
\cite{jen00,bir00}.  
To estimate the effect due to the finite width,
we construct the 3-D wave packet by superposing the eigenfunctions of 
the modes consistent with our phase-velocity filter \cite{bog97}. 
The FWHM of the energy distribution in the radial direction
for the wave packets of $N=7$ and 9 at the lower turning point is 
about $0.1 R_{\odot}$, which is greater than the separation 
between the lower turning points of two neighboring wave packets, 
about $0.02-0.03 R_{\odot}$. 
Thus the influence of magnetic fields at the base of the CZ on 
the wave packets of $N=7$ and 9 is not negligibly small compared with 
that on $N=8$.  This contradicts to the result of our measurements.
At this moment, we do not know how the combination of magnetic field 
perturbation and sound speed perturbation can help resolve this contradiction.

\acknowledgments

We thank T. L. Jr. Duvall and A. G. Kosovichev for helpful discussion
and comments.
We thank Jesper Schou for providing the frequency tables of MDI/SOHO.
SOHO is a project of international cooperation between ESA and NASA
This work utilizes data obtained by the Global Oscillation Network
Group (GONG) project, managed by the National Solar Observatory, which
is operated by AURA, Inc. under a cooperative agreement with the
National Science Foundation.  The data were acquired by instruments
operated by the Big Bear Solar Observatory, High Altitude Observatory,
Learmonth Solar Observatory, Udaipur Solar Observatory, Instituto de
Astrofisico de Canarias, and Cerro Tololo Interamerican Observatory.
We are grateful to the GONG Data Team for providing the software
package GRASP.
Authors were supported by NSC of ROC under grant NSC-90-2112-M-007-036.


\clearpage
%
\begin{figure}[t]
\epsscale{0.8}
\plotone{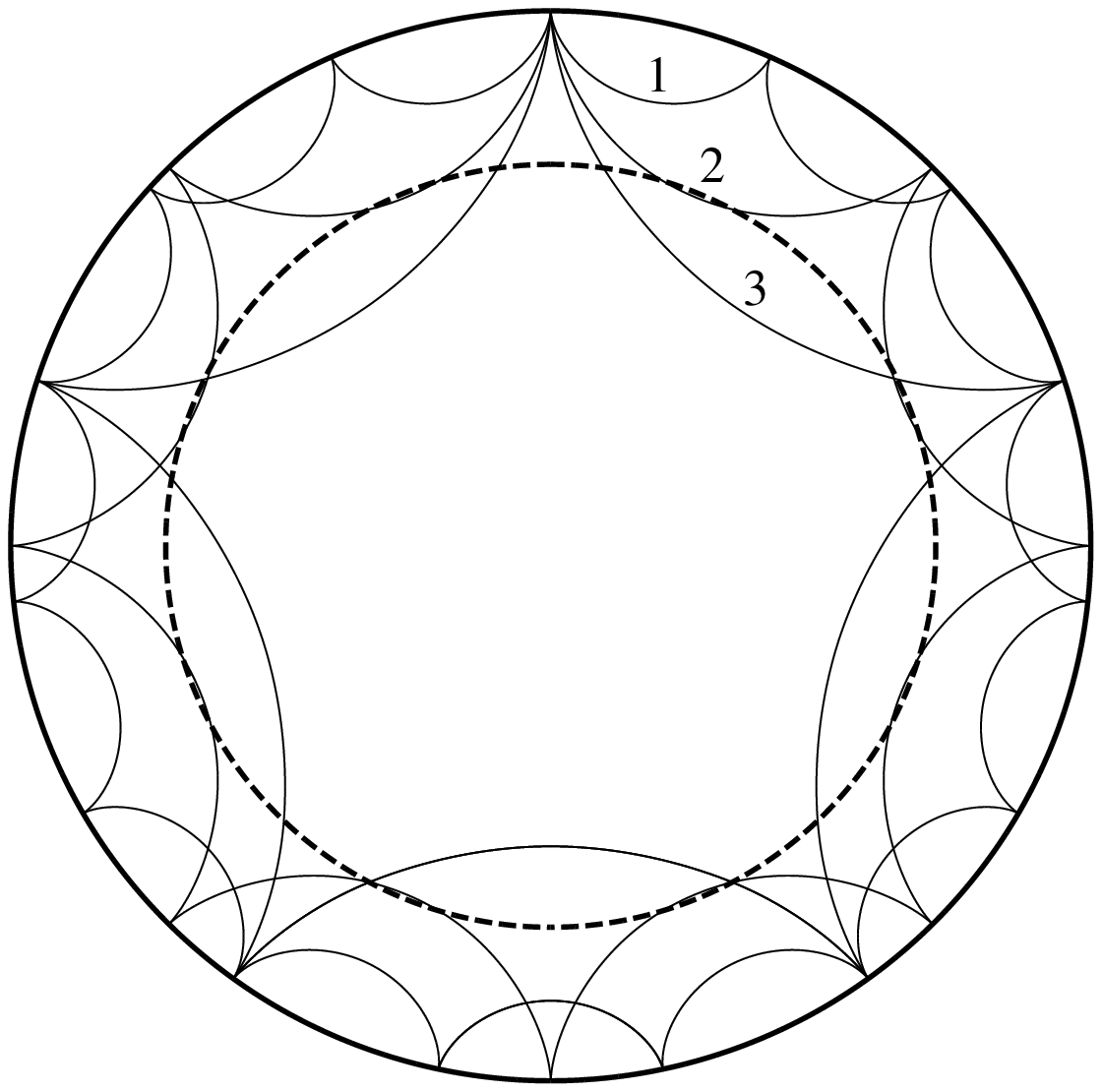}
\figcaption[f1.eps]{
Diagram showing ray paths of three different wave packets computed from
a standard solar model \cite{chr96} with the ray approximation \cite{dsi95}.  
The thick solid line is the solar surface, and 
the dashed line is the base of the CZ at $0.713 R_\odot$ \cite{bas97}. 
Ray 1, which takes 15 bounces to go around the Sun,
is not affected by the magnetic fields at the base of the CZ.
Ray 2, taking 8 bounces to go around the Sun,
has the lower turning point very close to the base of the convection zone.
Ray 3, taking 5 bounces to go around the Sun, can penetrate into the
radiative zone.  
\label{fig1}}
\end{figure}
%
\begin{figure}[t]
\epsscale{0.9}
\plotone{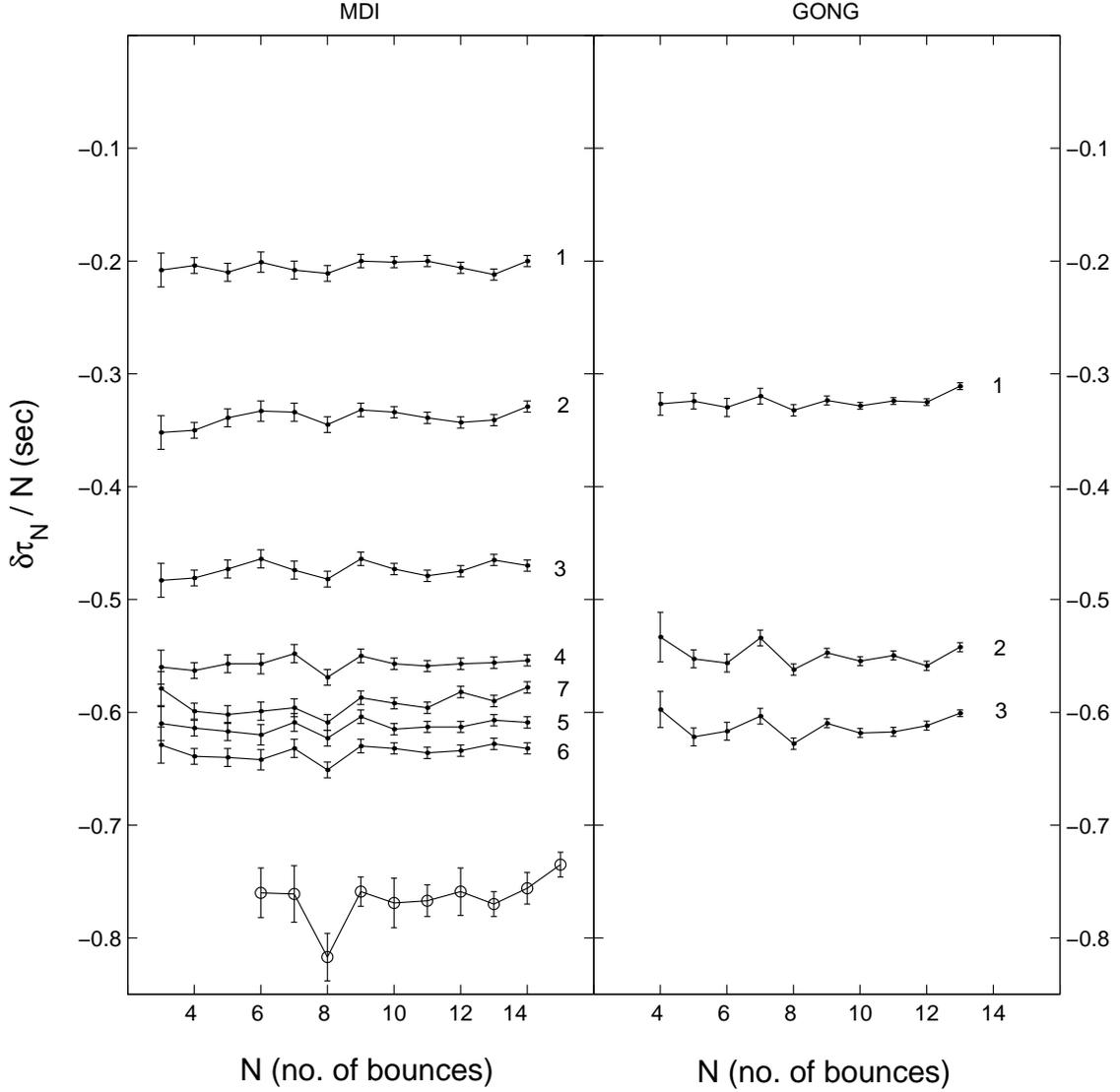}
\figcaption[f2.eps]{
Change in one-bounce travel time relative to solar minimum 
versus number of bounces $N$, which corresponds to different wave packets.  
The result from the multiple-bounce travel time analysis (MBTTA) is denoted
by the open circle in the left panel.  
The error bar is an estimate of fluctuation from 
averaging the travel time of different time series over solar minimum 
or maximum, corrected for variations of solar activity.  
The result from the power spectrum simulation analysis (PSSA) is denoted
by the filled circle.   
The left panel is computed from the MDI mode frequencies, and the right panel
from the GONG mode frequencies.
The sequence of the averaging periods is indicated by the number associated with
each curve.  
The range of each period is indicated by the horizontal bar
of each point in Figure 3. 
The error bar of PSSA is estimated from Monte Carlo simulations using 
the errors in mode frequencies. 
\label{fig2}}
\end{figure}
%
\begin{figure}[t]
\epsscale{0.9}
\plotone{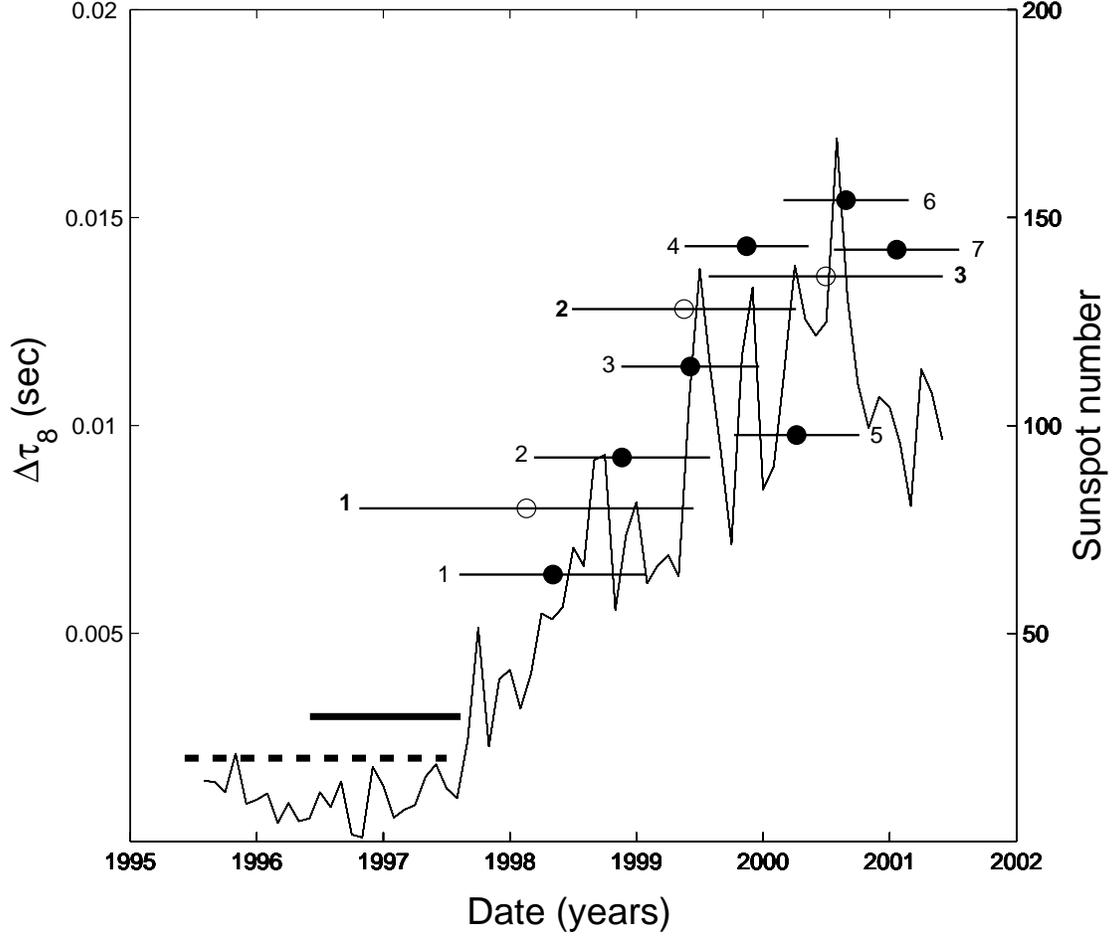}
\figcaption[f3.eps]{
Additional decrease in one-bounce travel time at $N=8$ relative to other $N$'s,
$-\Delta\tau_8$, from PSSA versus date.
The filled circles denote the MDI results, and the open circles the 
GONG results.
The horizontal bar associated with each point indicates the range of period 
used in averaging the mode frequencies. 
The number associated with each period is consistent with the 
number in Figure 2. 
The thick horizontal line indicates the range of solar minimum period 
used for MDI, and the dashed line for GONG.
The solid line is the sunspot number from the Greenwich Sunspot Data.
\label{fig3}}
\end{figure}

\end{document}